\documentclass[11pt,a4paper]{article}
\usepackage{amsfonts,latexsym,graphicx}

\newtheorem{thm}{Theorem}[section]
\newtheorem{lem}[thm]{Lemma}

\newtheorem{oldproof}{Proof}
\newenvironment{proof}[1][{}]{\begin{oldproof}}{\end{oldproof}}

\newcommand{\cL}{{\cal L}}
\newcommand{\1}{\mathbb{I}}
\newcommand{\U}{{\sf U}}
\newcommand{\R}{\mathbb{R}}
\newcommand{\C}{\mathbb{C}}
\newcommand{\T}{\mathbb{T}}

\newcommand{\Js}[1]{F_{+{#1}}}
\newcommand{\Jsb}[1]{F_{-{#1}}}
\newcommand{\Jsd}[1]{F_{\pm{#1}}}
\newcommand{\Jsa}[1]{F^{\dagger}_{-{#1}}}
\newcommand{\Jsab}[1]{F^{\dagger}_{+{#1}}}

\newcommand{\Jfd}[1]{{\sf F}_{\pm{#1}}}

\newcommand{\Jfad}[1]{{\sf F}^{\dagger}_{\pm{#1}}}

\newcommand{\M}[2]{M^{#1}_{+{#2}}}
\newcommand{\Mb}[2]{M^{#1}_{-{#2}}}
\newcommand{\Md}[2]{M^{#1}_{\pm{#2}}}
\newcommand{\Ma}[2]{M^{\dagger\,{#1}}_{-{#2}}}

\newcommand{\Sw}{\Psi}
\newcommand{\eSw}{\Xi}
\newcommand{\eSwa}{\Xi^{\dagger}}

\newcommand{\js}[1]{f_{+{#1}}}
\newcommand{\jf}[1]{{\sf f}_{+{#1}}}
\newcommand{\jfb}[1]{{\sf f}_{-{#1}}}
\newcommand{\jfd}[1]{{\sf f}_{\pm{#1}}}

\title{Inverse scattering for the matrix Schr\"{o}dinger operator and
Schr\"{o}dinger operator on graphs with general self-adjoint boundary
conditions}
\author{M.S. Harmer}
\date{}

\begin{document}

\maketitle

\begin{abstract}
Using a parameterisation of general self-adjoint boundary conditions in
terms of Lagrange planes we propose a scheme for factorising the matrix
Schr\"{o}dinger operator and hence construct a Darboux transformation an
interesting feature of which is that the matrix potential {\em and}
boundary conditions are altered under the transformation. \\
We present a solution of the inverse problem in the case of general
boundary conditions using a Marchenko equation and discusss the
specialisation to the case of graph with trivial compact part, ie.
diagonal matrix potential.
\end{abstract}

\section{The matrix Schr\"{o}dinger operator on the semi-axis}

We consider here the matrix Schr\"{o}dinger operator on the
semi-axis, ie. 
$$
\cL \equiv -\frac{d^2}{dx^2} + Q(x)
$$
on $L^2 (\R_+ , \C^n)$ where $\R_+ \equiv [0,\infty )$. The potential
matrix $Q(x)$ is assumed to be hermitian, absolutely integrable with
absolutely integrable first moment and continuous on the open
semi-axis \cite{Agr:Mar}. The matrix Schr\"{o}dinger operator, $\cL_0$,
defined on smooth functions vanishing at the origin and with compact
support, is a symmetric operator with deficiency indices $(n,n)$. Using
von Neumann extension theory \cite{Akh:Glz} we may parameterise all
self-adjoint extensions of this operator by unitary mappings between the
deficiency subspaces, ie.
$U\in\U (n)$. For practical purposes, however, it is more
convenient to describe the self-adjoint extensions in
terms of self-adjoint {\em boundary conditions} at the origin. \\
It is well known that the construction of self-adjoint extensions is
analogous to the description of Lagrange planes in a hermitian symplectic
space \cite{Kost:Sch}. In \cite{Kost:Sch} these Lagrange planes
are parameterised in terms of two
$n\times n$ matrices such that their product is hermitian. Really, in the 
case of a hermitian symplectic space which admits a canonical
basis\footnote{By our definition, a hermitian symplectic space for a
symmetric operator with non-equal deficiency indices does not admit a
canonical basis or Lagrange planes \cite{Har}.}, the Lagrange Grassmannian
is isomorphic to the unitary group
$\U (n)$ and we are able to explicitly parameterise the
Lagrange planes, and hence self-adjoint boundary conditions, in terms of
a unitary matrix---for details see \cite{Har}. In the case of the matrix
Schr\"{o}dinger operator on the semi-axis the self-adjoint boundary
conditions at the origin are given by
\begin{equation}\label{basicbc}
\frac{i}{2}(U^{\star}-\1 ) \left. \psi \right|_0 +
\frac{1}{2}(U^{\star} +\1 ) \left. \psi_{x} \right|_0 = 0 .
\end{equation}
Then the solution of the matrix Schr\"{o}dinger equation $\cL\eSw =
\lambda\eSw$ with boundary values 
\begin{equation}\label{eSwo}
\left. \eSw \right|_0 = \frac{1}{2}( U + \1 ) \equiv A ,\hspace{5mm} 
\left. \eSw_x \right|_0 = \frac{i}{2}( U -\1 ) \equiv B
\end{equation}
satisfies these boundary conditions. Using the Jost solutions, $\Jsd{}$,
solutions of the homogeneous equation $\cL\Jsd{} =\lambda\Jsd{}$, with 
asymptotic behaviour
$$ 
\lim_{x\rightarrow\infty}\Jsd{}(x,k) \sim e^{\pm ikx}\1 , \hspace*{5mm}
k=\sqrt{\lambda}
$$
and no prescribed behaviour at the origin, we can write 
\begin{equation}\label{eSwJs}
\eSw (x,k) = \Jsb{}(x,k)\Mb{}{}(k) + \Js{}(x,k)\M{}{}(k) .
\end{equation}
We define the scattering wave solution
\begin{eqnarray*}
\Sw (x,k) \equiv \eSw (x,k) \Mb{-1}{} = \Jsb{} + \Js{} S(k)
\end{eqnarray*}
where $S(k)$ is known as the scattering matrix. The
coefficients $\Md{}{}$ can be evaluated by taking the Wronskian of $\eSw$
and $\Js{}$ or $\Jsb{}$ \cite{Har}
\begin{equation}\label{sdmi}
\Md{}{} = \pm\frac{1}{2ik}\left[ \Jfad{} B - \Jfad{,x} A \right] .
\end{equation} 
where $\Jfd{}(k)\equiv\Jsd{}(0,k)$ are known as the Jost functions and 
$\mbox{}^{\dagger}$ is the involution
$Y^{\dagger} (x,k)\equiv Y^{\star}(x,\bar{k})$.
The Wronskian of $\eSwa$ and $\eSw$
$$
W\{\eSwa ,\eSw \} = \left. \left[ \eSwa \eSw_x - \eSwa_x \eSw \right] 
\right|_0 = A^{\star} B - B^{\star} A = 0 ,
$$
is always zero. Moreover, if we write
$\eSw$ in terms of the scattering wave solutions 
\begin{eqnarray*}
W\{\eSwa ,\eSw \} & = & \Ma{}{} W\{\Jsa{} + S^{\dagger} \Jsab{} ,\Jsb{} +
\Js{} S \} \Mb{}{} \\
& = & 2ik \Ma{}{} \left[ -\1 + S^{\dagger} S \right] \Mb{}{} = 0 .
\end{eqnarray*}
we see, since $S^{\dagger}=S^{\star}$ for $k\in\R$, that
the scattering matrix is unitary for real $k$. \\
If we diagonalise $U$, and use the well known asympototics of the Jost
functions \cite{Agr:Mar,Har} in the above expression for $\Md{}{}$, we
see that the scattering matrix has the following asymptotic behaviour:
\begin{lem}\label{umhu}
Given the self-adjoint operator $\cL$, with associated unitary matrix $U$
defining the boundary conditions of $\cL$, the scattering matrix of $\cL$
has the asymptotics\index{sma}
$$
\lim_{k\rightarrow\infty} S(k) \sim \hat{U}
$$
where $\hat{U}$ is a unitary {\em hermitian} matrix
$\hat{U}=\hat{U}^{\star}$ derived from $U$ by applying the map
$$
z\mapsto \left\{ \begin{array}{cl}
 1 & : z\in\T \setminus \{-1\} \\
-1 & : z = -1
\end{array}  \right.
$$
to the spectrum of $U$.
\end{lem}
Here $\T$ is the unit circle in $\C$. 
The matrix $\hat{U}$, since it is hermitian, defines projections
$$
{P} \equiv \frac{1}{2}(\1 +\hat{U}), \hspace{5mm}{P}^{\perp} \equiv
\frac{1}{2}(\1 -\hat{U}) , 
$$
which may be used to define a factorisation of the original operator
$\cL$. 
\begin{thm}\label{thDt}
Given the self-adjoint operator $\cL$ we can formally factorise it as
$$
\cL = D^{\star} D
$$
where
$$
D= i \left[ \frac{d}{dx} - V \right] 
$$
and the functions in the domain of $D$ satisfy the following boundary 
conditions at the origin,
$$
{P}^{\perp} \left.\psi\right|_0 = 0 .
$$
Furthermore, $V$ is a hermitian matrix which satisfies the Riccati
equation with the potential $Q(x)$ on the right hand side and has 
initial value satisfying 
\begin{equation}\label{Rebc}
{P} \left. V\right|_0 = - {P} H 
\end{equation}
where $H$ is a bounded hermitian matrix specified by the boundary
conditions of $\cL$.
\end{thm}
\begin{proof}
The fact that $V(x)$ satisfies the Riccati equation is well known. We
diagonalise the matrix $U$. Suppose that $l$ of the eigenvalues of $U$
are $-1$, then we choose the first
$l$ elements of the basis used in the diagonalisation to be the
eigenvectors with eigenvalue $-1$. In this basis the boundary conditions
of $\cL$ can be written
$$
\left[ \begin{array}{cc}
- i\1_{(l)} & 0 \\
0    & -\tan(\bar{\varphi}/2) \end{array} \right] 
\left. \psi\right|_0 +
\left[ \begin{array}{cc}
0 & 0 \\
0 & \1_{(n-l)} \end{array} \right]\left. \psi_x \right|_0 =0
$$
where $\psi$ is the boundary value (under the change of basis) and
$\varphi$ is the $(n-l)\times (n-l)$ diagonal matrix with entries which are
the eigenvalues of $U$ excluding the eigenvalues $-1$. This means that we can
write the boundary conditions for $\cL$ as
\begin{eqnarray}\label{Rebc0}
{P}^{\perp} \left.\psi\right|_0 & = & 0 \nonumber \\
{P} \left.\psi_x\right|_0 + {P} H \left.\psi\right|_0 & = & 0 
\end{eqnarray}
where $H$ is the hermitian matrix depending on the original boundary
conditions. Consider the operator $D^{\star}D$. From $D$ we get the 
boundary condition
$$
{P}^{\perp} \left.\psi\right|_0 =0
$$
and from $D^{\star}$ we get the boundary condition
$$
{P} \left.[ D \psi ]\right|_0 = 0
$$
which is
$$
{P} \left.\psi_x\right|_0 - {P} \left. V\right|_0
\left.\psi\right|_0 = 0 .
$$
So we see that as long as the initial value of $V$ satisfies equation
(\ref{Rebc}), $D^{\star}D$ has the required boundary conditions.
\end{proof}
It is well known that the Riccati equation can be linearised to the
Schr\"{o}dinger equation at zero energy, let us denote by
$\eSw_0$ these zero-energy solutions.
Then it is natural to ask how we can express the coefficient $V(x)$ in
terms of $\eSw_0$. The following theorem is proved in \cite{Har}, we
merely quote it here.
\begin{thm}
The hermitian matrix $V$ can be written
$$
V(x) = \eSw_{0,x} (x) \eSw^{-1}_0 (x)
$$
where $\eSw_0 (x)$ is the matrix of solutions of the Schr\"{o}dinger
equation at zero energy satisfying the boundary values of equation
(\ref{eSwo}) specified by the  unitary matrix ${U}_0$ and subject to:
\begin{enumerate}
\item The matrix $U_0$ satisfies
\begin{equation}\label{Dtpo}
P {U}_0 = P U . 
\end{equation}
\item The potential $Q(x)$ is continuous in some neighbourhood containing
the origin.
\end{enumerate}
\end{thm}
The inverse problem for the matrix Schr\"{o}dinger operator on the
semi-axis with Dirichlet boundary conditions (ie. vanishing of the
functions at the origin) is described in Agranovich and Marchenko \cite
{Agr:Mar}. In \cite{Har} the author extends Agranovich and Marchenko's
result to arbitrary self-adjoint boundary conditions at the origin; the
inverse problem may be reduced to a Marchenko equation\footnote{It is
also possible to reformulate it as a Riemann-Hilbert problem \cite{Har}.}
\begin{equation}\label{Me}
G(x+y) + K(x,y) + \int^{\infty}_{x} K(x,t) G(t+y)dt = 0 \hspace{5mm} x<y,
\end{equation}
where
$$
G(t) = \sum^{N}_{l=1} C^2_l e^{-\kappa_l t} + \frac{1}{2\pi}
\int^{\infty}_{-\infty} (S (k) - \hat{U}) e^{ikt} dk .
$$
Here $\lambda_l =-\kappa^2_l$ are the discrete eigenvalues and the $C_l$
are non-negative hermitian matrices known as the normalisation matrices.
The set $\{ S(k); \: \kappa_l, \: C_l, \: l=1,\ldots ,N \}$
is known as the scattering data. As this is a lengthy derivation we will
not repeat it here. If we are able to solve equation (\ref{Me}) for the
kernel of the transformation operator $K(x,y)$ we can recover the
potential matrix from the well known identity \cite{Agr:Mar}
$$
-2 \frac{d K(x,x)}{dx} = Q(x) .
$$
We can also recover the self-adjoint boundary conditions at the origin
from the inverse problem via
$$
U = \left[ \left.\Sw\right|_0 - \left.i \Sw_x\right|_0 \right] 
\left[ \left.\Sw\right|_0 + \left.i \Sw_x\right|_0 \right]^{-1} .
$$
which follows from the definition of the scattering wave solution plus
equation (\ref{eSwo}). Consequently, the solution of the inverse problem
allows us to recover not only the potential but also the self-adjoint
boundary conditions at the origin.

\section{The Schr\"{o}dinger operator on the graph with trivial compact
part}

The motivation for studying the matrix Schr\"{o}dinger operator is
that, {\em in the case of diagonal potential,} it may be identified with
the Schr\"{o}dinger operator on the non-compact graph with trivial compact
part---here we mean the graph consisting of $n$ semi-infinite rays with
the origin of each ray identified with the single vertex of the graph.
Although these are really two different operators, for the purposes of
the inverse problem they may be identified: each component of the vector
fuction on which the matrix Schr\"{o}dinger operator acts is identified
with the value of the function on one of the rays of the graph. \\
In this case the self-adjoint boundary conditions at the origin play a
crucial r\^{o}le; for instance if we have Neumann or
Dirichlet\footnote{Agranovich and Marchenko \cite{Agr:Mar} consider the
inverse problem for the matrix operator on the semi-axis, however, they
only consider Dirichlet boundary conditions which is not interesting in
the case of diagonal potential.} boundary conditions there is no
interaction between the rays and the graph decomposes into $n$ semi-axes
for which the solution of the inverse problem is well known.
The self-adjoint boundary conditions at the origin describe the
interaction between the rays and they may also be
thought of as inducing a `zero-range' potential at the origin
\cite{Har}. It is for these reasons that we consider the matrix
problem with general boundary conditions above. \\ 
For the problem on the graph---the diagonal matrix potential---the matrix
of Jost solutions is clearly a diagonal matrix and so too is the
kernel $K(x,y)$. As a result the Marchenko equation (\ref{Me}) has along
the diagonal $n$ scalar, independent (in the sense that each diagonal
entry of $K(x,y)$ appears only once) Marchenko equations. Consequently,
the inverse problem on the graph decouples to $n$ scalar inverse problems
which can always be solved using only the $n$ diagonal entries of the
scattering data\footnote{Actually, from equation (\ref{Me}) we need the
diagonal entries of $S(k)$ and $C^2_l$.}. Using this
scattering data we can recover the potential on the rays plus the
self-adjoint boundary conditions/zero-range potential at the origin. \\
If the self-adoint boundary conditions at the origin
are specified it may be possible to recover the potential on the rays
using a smaller set of scattering data. For Dirichlet or
Neumann boundary conditions there is no interaction between the rays
and we need all $n$ elements of the scattering data in order to recover
the potential. On the other hand, if there is an even number of rays
$n=2m$ and we have boundary conditions at the origin so that the graph
decomposes into $m$ copies of the whole real axis it is well known that
we can recover the potential using only $m$ reflection coefficients and
normalisation constants \cite{Nov0}. \\
Let us consider flux-conserved boundary conditions which are
defined by continuity at the origin
\begin{equation}\label{fcbc}
\left. \phi_1 \right|_0 = \left. \phi_2 \right|_0 = \cdots =
\left. \phi_n \right|_0 , \hspace*{8mm}
\sum^{n}_{i=1} \left. \phi^{\prime}_i \right|_0 = 0 
\end{equation}
plus conservation of flux. These are self-adjoint and we provide a brief
proof that in this case we need only $n-1$ of the diagonal elements of the
scattering data in order to recover the potential (for details see
\cite{Har}). We are able to show that the dispersion function is equal
to \cite{Har}
$$
\det (2ik\Mb{}{}(k)) \equiv M(k) = \frac{i^{n-1}}{n} \jf{,1} \jf{,2} \cdots \jf{,n} \sum^{n}_{i=1} 
\frac{\jf{,i}^{\prime}}{\jf{,i}}
$$
where $\jfd{,i}$ denotes the Jost function for the $i$-th ray of the graph
and $\jf{,i}^{\prime}(k)=\lim_{x\rightarrow 0}d\js{,i}(x,k)/dx$. The
scattering matrix has entries \cite{Har}, 
\begin{equation}\label{dsm}
S_{ij} (k) = \frac{2i^n  k\, \jf{,1} \jf{,2} \cdots \jf{,n}}
{n\, \jf{,i}\,\jf{,j}\, M(k)} - \delta_{ij}\frac{\jfb{,i}}{\jf{,i}} .
\end{equation}
Suppose we are given the first $n-1$ diagonal entries of the scattering
matrix for real $k$, the reflection coefficients, $R_j (k) \equiv S_{jj}
(k)$, plus the discrete eigenvalues $k_l = i\kappa_l$ and the first $n-1$
diagonal entries of the squares of the normalisation matrices, denoted
$b_{l,j}$, for each eigenvalue. Since the Marchenko equation
degenerates into
$n$ independent scalar equations on the rays we can use this data to
recover the potential and Jost solutions on the first $n-1$ rays.
Equation (\ref{dsm}) implies
$$
S_{ij} = \left[ R_i + \frac{\jfb{,i}}{\jf{,i}} \right] 
\frac{\jf{,i}}{\jf{,j}} \hspace*{5mm} \mbox{for}\; i\neq j .
$$
Consequently, we have enough information to recover the $(n-1)\times 
(n-1)$ minor of the scattering matrix formed by deleting the last column 
and row. Since $S(k)$ is unitary for $k\in\R$
we can recover the magnitudes of the remaining entries of the scattering
matrix. Let us consider any entry in the last column not on the diagonal
$$
\vert S_{in} \vert =  \left| \frac{2i^n k\, \jf{,1} \jf{,2} \cdots 
\jf{,n-1}}{n\, \jf{,i}\, M(k)} \right| \hspace*{5mm} i\ne n .
$$
As we have the Jost functions for $i=1,\ldots ,n-1$ we can solve this for
the magnitude of the dispersion function on the real axis.
Now $M(k)$ is analytic in the upper half-plane \cite{Har}, has known
magnitude on the real axis and known zeroes---the discrete
eigenvalues---and so we can recover $M(k)$ in the upper half-plane. 
To do this we consider the `normalised' dispersion function
$$
\hat{M}(k) = M(k)\frac{1}{i^n(k+i)} \prod^{N}_{l=1}\left[ 
\frac{k+i\kappa_l}{k-i\kappa_l} \right]^{m_{l}} .
$$
Here $i\kappa_l$ are the zeroes and 
$m_l$ the orders of the zeroes of $M(k)$. In \cite{Har} we describe in
detail how, using the $b_{l,i}$, to find the orders $m_l$  and also
show there that, subject to the absence of vitual levels, $\hat{M}(k)$ is
bounded and non-zero on the closed real axis with asymptotic
$$
\lim_{\vert k\vert\rightarrow\infty}\hat{M}(k) = 1 .
$$
Furthermore, by definition $\hat{M}(k)$ is analytic in the upper
half-plane with no zeroes there. By a simple application of the Cauchy
integral formula and the Plemelje formula (see 
\cite{Nov0} for a similar calculation in the scalar case)
\begin{eqnarray*}
\arg M(k) & = & \frac{1}{i} \left[ \ln i^n (k+i) + 
\sum^N_{l=1} m_l \ln \frac{k-i\kappa_l}{k+i\kappa_l} \right] - \\
& & \mbox{} -\frac{1}{\pi}\int^{\infty}_{-\infty} 
\frac{\ln\left[ \vert M (k^{\prime})\vert
/ \vert k^{\prime}+i \vert \right]} {k^{\prime}-k} dk^{\prime}
\end{eqnarray*}
where we take the principle value of the integral.
Consequently we recover $M(k)$. From $M(k)$ we can use equation
(\ref{dsm}) to recover $S_{in}$ for 
$i\neq n$. We now have enough information to recover the potential on the
last ray: from the scattering matrix we can find $\jf{,n}$ and
from $M(k)$ we can find $\jf{,n}^{\prime}$. These two functions are enough
to recover the potential on the ray as they provide the scattering data
for the Schr\"{o}dinger operator on the semi-axis \cite{Agr:Mar} (with
Dirichlet boundary conditions in the cited text). \\

\section{Acknowledgements}

The author would like to thank Prof B.S. Pavlov for his advice and
many useful conversations.


\begin{thebibliography}{1}

\bibitem{Agr:Mar}
Z.~S. Agranovich and V.~A. Marchenko.
\newblock {\em The Inverse Problem of Scattering Theory}.
\newblock Gordon and Breach, New York, 1963.

\bibitem{Akh:Glz}
N.~I. Akhiezer and I.~M. Glazman.
\newblock {\em Theory of Linear Operators in Hilbert Space}.
\newblock Frederick Ungar Publishing, New York, 1966.

\bibitem{Har}
M.~Harmer.
\newblock {\em The Matrix {Schr\"{o}dinger} Operator and {Schr\"{o}dinger}
  Operator on Graphs}.
\newblock PhD thesis, University of Auckland, 2000.

\bibitem{Kost:Sch}
V.~Kostrykin and R.~Schrader.
\newblock Kirchhoff's rule for quantum wires.
\newblock {\em J. Phys A: Math. Gen.}, 32:595--630, 1999.

\bibitem{Nov0}
S.~P. Novikov, S.~V. Manakov, L.~P. Pitaevskii, and V.~E. Zakharov.
\newblock {\em Theory of Solitons: The Inverse Scattering Method}.
\newblock Consultants Bureau, New York, 1984.

\end{thebibliography}
\end{document}